\documentclass[doublecol,A4paper]{epl2}

\usepackage{color}

\title{Single origin of the nodal and antinodal gaps in cuprates}

\author{Yves Noat\inst{1,*} \and Alain Mauger\inst{2} \and William Sacks\inst{2}}

\shortauthor{Y. Noat, A. Mauger \& W. Sacks}

\institute{\inst{1} Institut des Nanosciences de Paris (INSP), UMR 7588,\\

\inst{2} Institut de Min\'{e}ralogie, de Physique des\\
Mat\'{e}riaux, et de Cosmochimie (IMPMC), UMR 7590,\\

Sorbonne Universit\'{e}s, \\
4 place Jussieu, 75252 Paris Cedex 05, France \\

\inst{*}Corresponding author\,: yves.noat@insp.jussieu.fr}


\abstract{Recent angle-resolved photoemission electron spectroscopy
(ARPES) experiments demonstrate that the momentum dependence of the
spectral gap in underdoped cuprates does not follow a pure $d$-wave
form [H. Anzai et a., Nat. Comm. {\bf 4}, 1815 (2013)]. This
deviation is highly controversial. It has often been interpretated
as a proof of the non-superconducting origin of the antinodal gap in
the underdoped regime. In this article, we show that the measured
angular dependence of the spectral gap can be explained by the basic
nature of pairs in high-T$_c$ cuprates. Hole pairs, or {\it
pairons}, form as a result of the local antiferromagnetic
environment on the scale $\xi_{AF}$, the magnetic coherence length.
The spatial extension of the pairon wavefunction beyond first
nearest neighbours gives rise to the anomalous angular dependence of
the gap, in quantitative agreement with experiments. This simple
interpretation strongly indicates a common origin of the nodal and
antinodal gaps.}

\begin{document}

\maketitle

\vskip 2mm {\it Introduction: Are there two energy scales in superconducting cuprates?} \vskip 2mm


One of the most challenging problems of high-T$_c$ cuprates is the
interpretation of the spectral gap. Contrary to conventional
superconductors where the spectral gap vanishes at the critical
temperature, as described by the Bardeen-Cooper-Schrieffer (BCS)
theory \cite{PR_BCS1957}, in cuprates a gap persisting at the
critical temperature T$_c$, the so-called pseudogap, is directly
observed in angle resolved photoemission spectroscopy (ARPES)
\cite{Nat_Ding1996,Sci_Loeser_1996} or scanning tunneling
spectroscopy \cite{PRL_renner1998_T}. Whether or not the pseudogap
is linked to superconductivity remains a fundamental question.

More recently, low-temperature experiments revealed a non-trivial
angular dependence of the gap as a function of doping. While in
overdoped samples the gap follows a strictly $d$-wave behavior, a
clear deviation from $d$-wave is observed on the underdoped side of
the phase diagram (see \cite{JPhysSocJap_Yoshida2012} and Refs.
therein). This so called two-gap behavior was first revealed by
electronic Raman scattering \cite{NatPhys_LeTacon2006} and
photoemission \cite{Sci_Tanaka2006} and later confirmed by tunneling
spectroscopy \cite{Sci_Pushp2009}. It is most often interpreted as
the result of the competition between superconductivity and the
phase responsible for the pseudogap
\cite{Sci_Tanaka2006,Nat_Kondo2009,PRL_Yoshida2009,PNAS_Vishik2012}.

As first noted by Kondo et al. \cite{PRL_Kondo2007} and Terashima et al. \cite{PRL_Terashima2007}, the shape of $\Delta(\theta)$ can be properly described by including an additionnal harmonic term to the usual $d$-wave dependence:
\begin{equation}
\Delta(\theta)=\Delta_0 \left[B\cos(2\theta)+(1-B)\cos(6\theta)\right]
\label{Delta_theta_Ando}
\end{equation}
An analogous formula was later used by Anzai et al.\,\cite{Natcom_Anzai2013} who carefully studied the angular dependence of the gap as a function of doping. The fit of the ARPES data can be used to determine the doping dependence of the nodal and antinodal gaps. By this procedure, Anzai et al. found a dome shape for the nodal gap, which was then interpreted as the order parameter \cite{Natcom_Anzai2013}, suggesting that the superconding (SC) state is restricted to the near nodal region in $k$-space (i.e. in the Fermi arc region, as in Ref.\,\cite{PRL_Yoshida2009}).

A separate analysis performed by Vishik et
al.\,\cite{PNAS_Vishik2012} led to a completely different
conclusion, namely that the nodal gap is almost doping independent
in the underdoped regime while it is almost equal to the antinodal
value in the overdoped regime. The crossover between the two
behaviors occurs for a doping value $p\sim$ 0.18 (i.e. slightly
higher than the optimum doping value $p\sim$ 0.16). This discrepancy
between the two analyses clearly indicates that a more careful
attention should be paid to the method used to infer the nodal gap
value. From an overview of the litterature on this question
\cite{JPhysSocJap_Yoshida2012}, one concludes that the connection of
the nodal and antinodal gaps to the SC state remains an important
unsolved issue for understanding high-T$_c$ superconductivity.

In this paper, we reconcile the contradictory experimental analysis and give a simple explanation for the angular dependence of the gap in the whole doping range, from underdoped to overdoped regimes. Our analysis demonstrates that the latter is essentially governed by the doping dependence of the antiferromagnetic coherence length $\xi_{AF}$, which determines the size of pairs forming the SC condensate.

\vskip 2mm {\it Angular dependence of the gap} \vskip 2mm

There is an overall agreement that, at least in the overdoped regime, the spectral gap as a function of angle has a simple $d$-wave form:
\begin{equation}
\Delta(\vec{k})=\Delta_p\left[\cos(k_xa)-\cos(k_ya)\right]
\label{Delta_dwave}
\end{equation}
At the Fermi surface ($k=k_F$), this formula can be well approximated by
\begin{equation}
\Delta(\theta)\simeq\Delta_p \cos(2\theta)
\label{Delta_dwave_theta}
\end{equation}
The angle in the above formula is taken from the ($\pi$,$\pi$)
corner of the first Brillouin zone. With these coordinates, the
antinodal direction ($\pi$,0) corresponds to $\theta=0$ while the
nodal direction ($k_x=k_y$) corresponds to $\theta=\frac{\pi}{4}$.

\begin{figure}
\includegraphics[width=8.4 cm]{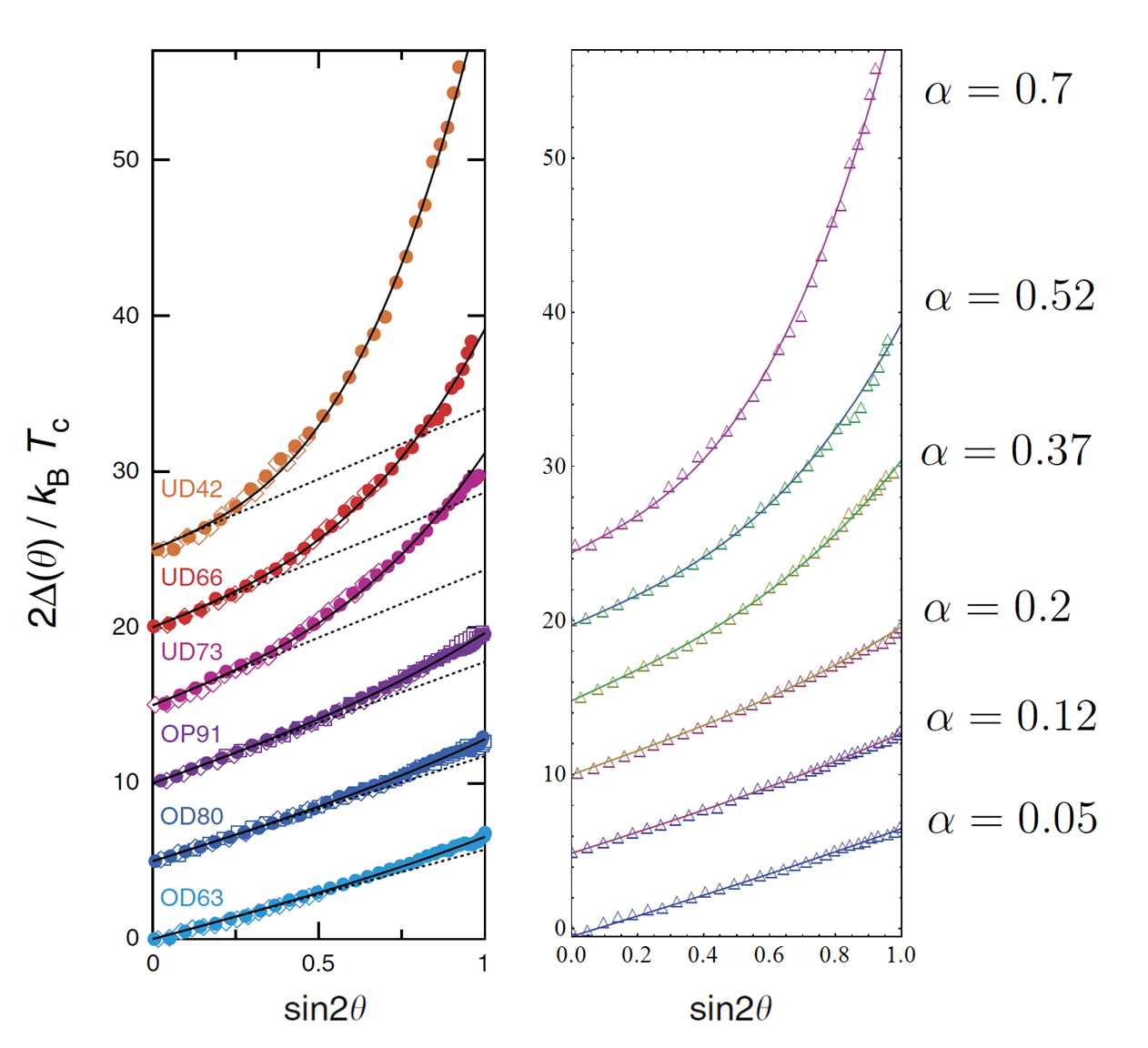}
\caption{(Color online) Left panel: Gap as a function of
$\sin(\theta)$ from the work of Anzai et
al.\,\cite{Natcom_Anzai2013}; Right panel: Fit of the data
$\Delta(\theta)$ using formula \ref{formula_WS}. The values of
$\alpha$ deduced from the fits are indicated on the right side for
each doping value. (Note that in ref.\,\cite{Natcom_Anzai2013} the angle
is measured from the node.)}\label{Fig_Ino}
\end{figure}

\begin{figure}
\includegraphics[width=8.4 cm]{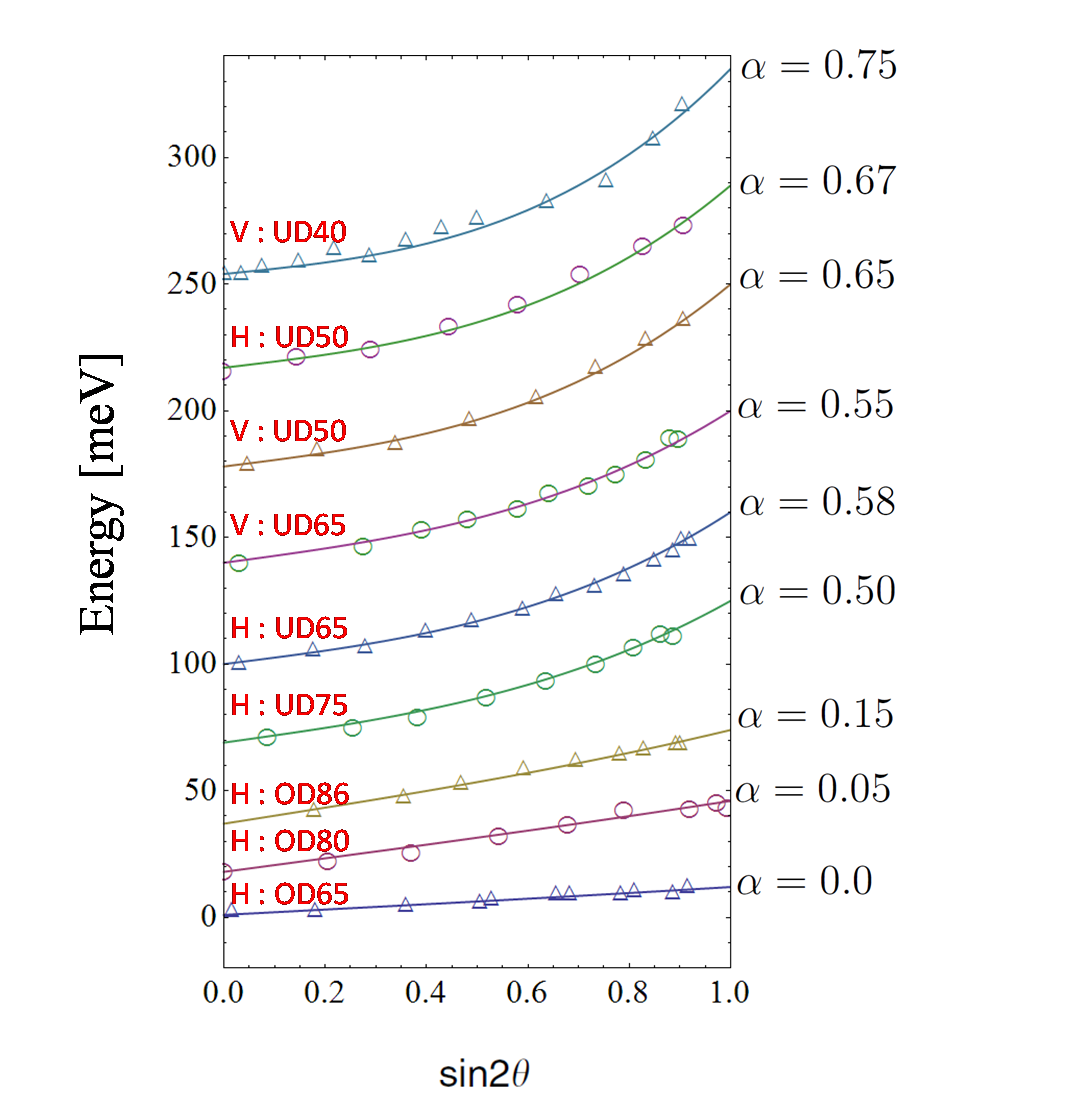}
\caption{(Color online) Angular dependence of the gap from Vishik et al.\,\cite{PNAS_Vishik2012} (spectra labeled with letter `V') and Hashimoto et al.\,\cite{NatPhys_Hashimoto2014} (spectra labeled with letter `H'); Right panel: Fit of $\Delta(\theta)$ with formula \ref{formula_WS}. The values of $\alpha$ deduced from the fits are indicated on the right side for each doping value.} \label{Fig_Vishik_Hashi}
\end{figure}

Deviations from this standard form have been observed in the underdoped regime by several groups \cite{PRL_Kondo2007,PRL_Terashima2007,PNAS_Vishik2012,Natcom_Anzai2013,NatPhys_Hashimoto2014}. Such deviations, as in Fig.\,\ref{Fig_Ino} and \ref{Fig_Vishik_Hashi}, occur if the doping is lower than some value, which varies from experiment to experiment (see \cite{JPhysSocJap_Yoshida2012} for a review). The latter is generally higher than the optimum doping value \cite{JPhysSocJap_Yoshida2012}.

Although the angular dependence of $\Delta(\theta)$ includes new harmonics as mentionned above, in the underdoped regime the overall shape still has a $d$-wave character. We find it useful to reformulate Eq. \ref{Delta_theta_Ando}, in an alternate algebraic form:
\begin{equation}
\Delta(\theta)=\Delta_p\left[1-\alpha \sin^2(2\theta)\right]\cos(2\theta)
\label{formula_WS}
\end{equation}
In this form, the gap at the antinode is by definition $\Delta(\theta=0)=\Delta_p$ while the so-called nodal gap is defined by:
 \begin{equation}
\Delta_N=-\frac{1}{2}\frac{d\Delta(\theta)}{d\theta}|_{\theta=\frac{\pi}{4}}=\Delta_p(1-\alpha).
\label{Nodal_gap}
\end{equation}
Note that while the common name often given to $\Delta_N$ is the nodal `gap', it is in fact {\it the slope} of $\Delta(\theta)$ at the node. Thus, $\alpha$ appears as the key parameter giving rise to the difference between the nodal and the antinodal gaps.

We have fit the experimental spectra from Anzai et
al.\,\cite{Natcom_Anzai2013} for different doping values. As shown
in Fig. \ref{Fig_Ino} the agreement between Eq.\ref{formula_WS} and
the experiment is remarkable. While in the overdoped regime, the
shape of $\Delta(\theta)$ is perfectly $d$-wave, a clear deviation
is observed in the underdoped side of the phase diagram. This
deviation grows as the doping is lowered. We have also analysed data
from Vishik et al.\,\cite{PNAS_Vishik2012} as well as the one
reported by Hashimoto et al.\,\cite{NatPhys_Hashimoto2014}. As shown
in Fig. \ref{Fig_Vishik_Hashi}, Eq. \ref{formula_WS} also perfectly
reproduces the angular dependence of the gap they have observed in a
wide doping range.

We now come to the determination of the slope of $\Delta(\theta)$,
i.e. the nodal gap. In ref.\,\cite{PNAS_Vishik2012}, the latter was derived from the linear fit of the gap value near the node. This method, however, is not correct, because the curvature of the experimental curve does not vanish at any value of $\theta$ in the underdoped regime, so that the slope found by assuming {\it a priori} a linear behavior near the node depends on both the number of experimental points and the extention in $\theta$ chosen to make such a fit.

Contrary to the procedure used in
\cite{PNAS_Vishik2012}, where the slope was tentatively estimated
from the experimental points near the node, we calculate the nodal
gap from the fit of $\Delta(\theta)$ using Eq. \ref{Nodal_gap}. This
is a fundamental point in the analysis. Indeed, it should be
stressed that in order to deduce precisely the slope of
$\Delta(\theta)$ at the node, given by Eq. \ref{formula_WS}, one
needs to properly fit the whole curve from the node to the antinode.
This difference explains the discrepancy in the determination of
$\Delta_N$ reported in \cite{PNAS_Vishik2012} and
\cite{Natcom_Anzai2013}.

\begin{figure}
\centering \vbox to 7.8 cm{
\includegraphics[width=8.5 cm]{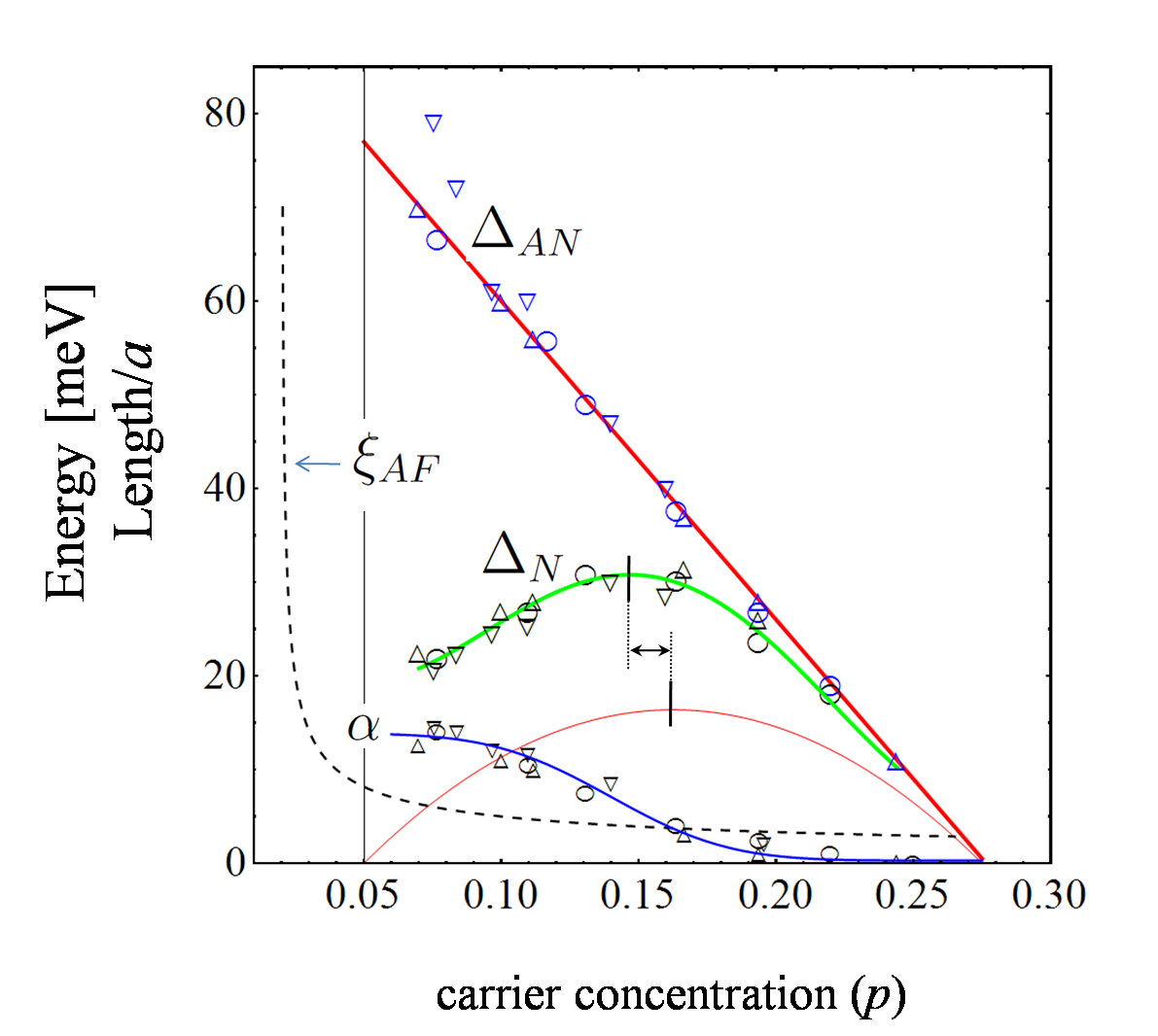}}
\caption{(color online) Phase diagram deduced from the fits from experimental data taken from Anzai et al.\,\cite{Natcom_Anzai2013} (circles), Vishik et al.\,\cite{PNAS_Vishik2012} (inverted triangles) and Hashimoto et al.\,\cite{NatPhys_Hashimoto2014} (triangles). The nodal and antinodal gaps and the parameter $\alpha$ deduced from the fits as a function of doping.}\label{Fig_Phase_diag}
\end{figure}

The nodal gap inferred from our analysis is plotted
in Fig.\,\ref{Fig_Phase_diag} which proves the consistency of the
data of both Refs. \cite{Natcom_Anzai2013} and
\cite{PNAS_Vishik2012}. Although the nodal gap has a dome shape, it
is nevertheless clear that it does not follow T$_c$. The maximum of
$\Delta_N(p)$ is shifted with respect to that of T$_c(p)$. Moreover
$\Delta_N(p)$ does not vanish for $p=p_{min}$ where the
superconducting phase first appears. This finding seems to discard the hypothesis that the nodal gap is directly the
order parameter, contrary to the conclusion raised in \cite{Natcom_Anzai2013}.

In addition, the doping dependence of the parameter $\alpha$ which characterizes the deviation from $d$-wave, closely follows the one deduced from the data of Anzai et al.\,\cite{Natcom_Anzai2013}, which gives good confidence in our analysis.

The present approach thus reconciles ARPES experiments where contradictory interpretations were given. We now turn to the physical origin of the anomalous angular dependence.

\vskip 2mm {\it Pairons: hole pairs in a short-range antiferromagnetic background} \vskip 2mm

As demonstrated by pionieering works in the early nineties in the context of the Hubbard or $t-J$ hamiltonians, two holes in an antiferromagnetic background can form a bound state for a sufficiently large $t/J$ ratio \cite{PRB_Kaxiras1988,PRB_Riera1989,PRB_Bonca1989,PRB_Hasegawa1989}, with the expected $d$-wave symmetry \cite{PRB_Poilblanc1994}. These works strongly suggest the existence of hole pairs in an AF background, although the description in terms of strongly correlatted electrons might fail in the overdoped side of the phase diagram \cite{PRL_Hardy2009}.

We have extended this concept to realistic systems
\cite{EPL_Sacks2017} where the key point is the
small but finite antiferromagnetic coherence length. Indeed, in
cuprates, the long range antiferromagnetic state is broken by hole doping at $p=p_{AF}$, giving rise to a superconducting state with short-range AF correlations for $p>p_{min}$ \cite{PRB_Takagi1989}. The superconducting phase emerges from the metallic phase in the doping range $p_{min}<p<p_{max}$.

In recent articles, we have proposed that high-T$_c$
superconductivity can be explained in terms of the formation of hole
pairs, or {\it pairons}, in their local antiferromagnetic
environment \cite{EPL_Sacks2017}, on the scale of $\xi_{AF}$.
In \cite{EPL_Sacks2017} each pairon is associated
with an AF cell of area $\xi_{AF}^2$ which naturally explains the
linear behavior of the antinodal gap energy with doping.

Two different temperatures can be distinguished corresponding
respectively to the formation of the pairs and their condensation,
both being linked to a single energy scale, $J$, the exchange
energy. Pairons form at a temperature $T^*$ higher than the critical
temperature and undergo a Bose-like condensation at T$_c$ as a
result of repulsive pair-pair interactions \cite{SciTech_Sacks2015}.

The pairon model is supported by several experimental facts, in particular by
\vspace{0,2cm}

(i) the presence of a low-temperature energy gap within the vortex core where coherence is lost \cite{PRL_renner1998_B}, which strongly suggests the presence of preformed pairs,

ii) the concomitant disappearance of superconductivity and antiferromagnetic correlations as a function of doping (for $p=p_{max}$) \cite{PhysicaC_Yoshizaki1990},

iii) the scaling of the critical temperature with the exchange energy $T_c\propto J$ \cite{PRB_Ofer2006}, which is a direct consequence of the nature of pairons and of the corresponding phase diagram \cite{EPL_Sacks2017}.

The nature of pairons is revealed in the excitation spectrum as a function of temperature. Indeed, two well-defined signatures are observed in ARPES measurements: a Fermi arc near the node at T$_c$ and below \cite{Nat_Norman1998} and a gap without coherence peaks, the pseudogap, in the antinode above T$_c$ \cite{Nat_Ding1996,Sci_Loeser_1996}. The latter display the characteristic Bogoliubov quasiparticle dispersion indicating preformed pairs \cite{PRL_Kanigel2008,EPL_Shi2009}. In the pairon model, the Fermi arc is the result of fermion excitations, due to the composite fermion character of pairons, while the pseudogap is a manifestation of an incoherent state of excited pairons, absent in a conventional BCS superconductor \cite{Jphys_Sacks2018}.


\vskip 2mm {\it Origin of the angular dependence of the gap} \vskip 2mm

\begin{figure}
\centering \vbox to 5. cm{
\includegraphics[width=8.5 cm]{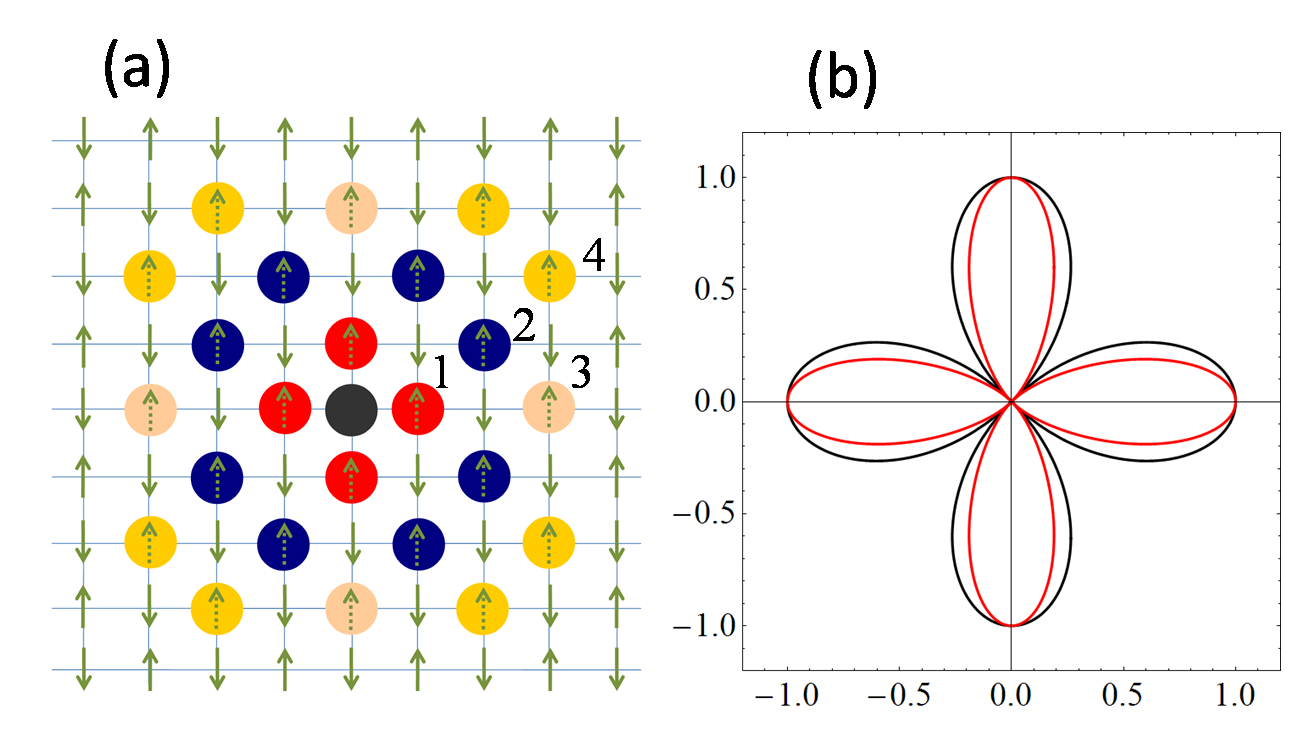}}
\caption{(color online) a) Illustration of an hole pair in its antiferromagnetic environment. The red/blue/pink/yellow dots correspond respectively to first, second, third and fourth nearest neighbours for the hole pairs, one hole (black hole) being located in the center. b) Pure $d$-wave order parameter ($\alpha=$0, black curve), compared to gap including higher harmonics term ($\alpha=$0.7, red curve). As can be noted, the gap is reduced near the node compared to the pure $d$-wave gap (black curve).}\label{Fig_Pairon}
\end{figure}

The pairon model also provides a physical
explanation for the observed gap angular dependence. Indeed, to
understand the shape of $\Delta(\theta)$, one needs to consider how
the gap is modified at low doping as the size of antiferromagnetic
correlations increases. For this purpose, we start from the
expression of the hole pair wavefunction in relative coordinates. In
our previous work \cite{Jphys_Sacks2018}, only first nearest
neighbours were considered (red dots in Fig. \ref{Fig_Pairon}):
\begin{eqnarray}
 \psi_{pair}^{(1)}(\vec{r})& = & \frac{1}{\sqrt{4}}\left[\varphi(\vec{r}-a\hat{x})+\varphi(\vec{r}+a\hat{x})\right. \label{Psi_first neighbours}\\
&-& \left.\varphi(\vec{r}-a\hat{y})-\varphi(\vec{r}+a\hat{y})\right] \nonumber
\end{eqnarray}
where $a$ is the lattice parameter, and
$\varphi(\vec{r})={e^{-\frac{{r}^2}{2b^2}}}/{\sqrt{2\pi b^2}}$. The
parameter $b$ fixes the spatial extension of the pair wavefunction
in relative coordinates. The overlap between the various amplitudes
in Eq. \ref{Psi_first neighbours} has been neglected. Note that the
sign in the above formula originates from the antiferromagnetic
symmetry which imposes that the wavefunction must vanish in the
lattice diagonal (corresponding to the nodal direction in
$k$-space).

It follows that the associated gap is expressed in the standard
$d$-wave form (Eq. \ref{Delta_dwave}) \cite{Jphys_Sacks2018}:
\begin{equation}
\Delta(k)=\tilde{\varphi}(k_F)\,A_1\,g_1(k_x,k_y),
\end{equation}
where $\tilde{\varphi}(k_F)$ is the Fourier transform of $\varphi(\vec{r})$ taken at $k=k_F$, $A_1$ is a constant and
\begin{equation}
g_1(k_x,k_y)=\cos(k_xa)-\cos(k_ya)
\end{equation}


As the doping is lowered, the size of the pairons, of the order of $\xi_{AF}$, increases. The delocalized wavefunction must then include the contribution of next nearest neighbours (blue dots in Fig. \ref{Fig_Pairon}):
\begin{equation}
\psi_{pair}(\vec{r})=
\sum_i \psi_{pair}^{(i)}(\vec{r})
\label{Phi_allterms}
\end{equation}
with $\psi_{pair}^{(i)}(\vec{r})\propto\sum_{\vec{r_j}}c_j\varphi(\vec{r}-\vec{r_j})$; the sum runs over all the i$^{th}$ nearest neighbours and $c_j=\pm$1 determines the sign of each term and is imposed by the AF symmetry, as explained below.
The second neighbour correction is expressed as
\begin{eqnarray}
&&\psi_{pair}^{(2)}(\vec{r})\propto \nonumber\\
&&-\left[\varphi(\vec{r}-2a\hat{x}-2a\hat{y})+ \varphi(\vec{r}-a\hat{x}-2a\hat{y})\right.\nonumber\\
&&+\varphi(\vec{r}+a\hat{x}-2a\hat{y})-\varphi(\vec{r}+2a\hat{x}-a\hat{y})\label{Phi_secondterm}\\
&&-\varphi(\vec{r}+2a\hat{x}+a\hat{y})+\varphi(\vec{r}+a\hat{x}+2a\hat{y})\nonumber\\
&&+\left.\varphi(\vec{r}-a\hat{x}+2a\hat{y})-\varphi(\vec{r}-2a\hat{x}+a\hat{y})\right.]
\nonumber
\end{eqnarray}

The Fourier transform of the wavefunction $\psi_{pair}(\vec{r})$
(Eq. \ref{Phi_allterms}) then gives rise to additional terms. As
described in a previous article \cite{Jphys_Sacks2018}, the pairon
wavefunction can then be expressed as a superposition of Cooper
pairs in $k$-space. Extending the calculation derived in
\cite{Jphys_Sacks2018}, we obtain the gap function:
\begin{equation}
\Delta(\vec{k})=\tilde{\varphi}(k_F)\sum_{i}A_i\,g_i(k_x,k_y)
\label{Delta_general form}
\end{equation}
where $\tilde{\varphi}(k_F)\,A_i\,g_i(k_x,k_y)$ is the contribution to the gap amplitude of the i$^{th}$ nearest neighbours. The first term has the standard $d$-wave form, while the second nearest neighbour contribution, deduced from Eq. \ref{Phi_secondterm}, is expressed as:
\begin{eqnarray}
g_2(\vec{k})& = & \left[-\cos(2k_xa+k_ya)+\cos(k_xa+2k_ya)\right. \\
&+& \left.\cos(k_xa-2k_ya)-\cos(2k_xa-k_ya)\right] \nonumber
\label{g_seconds_voisins}
\end{eqnarray}
There are two possible signs for the second neighbour correction to
the wavefunction, $g_2(\vec{k})$. The positive sign in front of the
bracket was chosen in order to properly describe the experimental
results. It is compatible with the AF symmetry, introducing new
nodal lines in the second neighbour correction, in agreement with
the higher angular harmonics of Eq.{\ref{formula_WS}}. This effect
{\it reduces} the nodal gap compared to the antinodal value (see
Fig. \ref{Fig_Pairon}b).

\vskip 2mm {\it Discussion} \vskip 2mm

The previous expression accounts quantitatively for the angular
dependence of the gap. The reason is that, for any choice of $A_1$
and $A_2$, the above formula is strictly equivalent to formula
\ref{formula_WS} used for the fits (i.e. there is an equivalence
between the parameter $\alpha$ used in Eq. \ref{formula_WS} and
$A_2/A_1$). The correspondence between the parameter using formula
\ref{formula_WS} and \ref{Delta_general form} restricted to second
neighbours can be found in table \ref{param_table}.
It is relatively straightforward to express the
coefficients $A_i$ reflecting the weight of the i$^{th}$ neighbour
in the hole pair wavefunction. We have numerically evaluated $A_i$
for $i=3$ and $4$ which turn out to be negligible so that only
second neighbours contribute to the angular dependence of the gap at
low doping.

\begin{table*}
\centering
\begin{tabular}{c|c|c|c|c|c|c|c|c}
$p$     & $T_c$ & $\frac{2\Delta}{kT_c}$ & $\Delta_p$ & $\Delta_N$ & $\alpha$ & $\eta$     & $\frac{A_2}{A1}$  & $\xi_{AF}$ \\
  & (K) &  & (meV) & (meV) &  &  &  & ($a$ units) \\ \hline
0.25  & 37.5 & 5.6                          & 9                        & 9                        & 0                     & 0     & 0     & 2.9                      \\
0.22  & 63   & 7                            & 19                       & 19                       & 0.05                  & 0.015  & 0.01  & 3.16                     \\
0.194 & 80   & 7.8                          & 26.9                     & 23.6                     & 0.12                  & 0.051 & 0.036 & 3.4                      \\
0.164 & 91   & 9.6                          & 37.7                     & 30.16                    & 0.2                   & 0.082 & 0.06 & 3.7                      \\
0.131 & 73   & 15.6                         & 49                       & 30.87                    & 0.37                  & 0.185 & 0.13 & 4.2                      \\
0.117 & 66   & 19.6                         & 55.8                     & 26.8                     & 0.52                  & 0.275  & 0.2 & 4.5                      \\
0.077 & 42   & 36.8                         & 66.6                     & 21.9                     & 0.7                   & 0.364 & 0.295 & 5.9
\end{tabular}
\caption{Experimental parameters ($T_c$, $\Delta_p$) from Anzai et al \cite{Natcom_Anzai2013} and parameters deduced from the fits (the nodal gap $\Delta_N$, $\alpha$, $\eta$, $\frac{A_2}{A_1}$) as a function of the doping value $p$. We also indicate the calculated value of $\xi_{AF}(p)$ ($a$ units) obtained using Eq. \ref{xi_AF mod} with $p_{AF}=$0.02 for each doping value.}
\label{param_table}
\end{table*}

The second neighbour correction to the gap must be directly related
to the extension of the pairon wavefunction, i.e. to the size of the
AF correlations. Let us rewrite Eq. \ref{Delta_general form} in the
equivalent form:
\begin{equation}
\Delta(\vec{k})=\tilde{\varphi}(k_F)\,A_1\left[g_1(k_x,k_y)+\frac{A_2}{A_1}g_2(k_x,k_y)\right]
\label{2nd_neighbor}
\end{equation}
Assuming that $\frac{A_2}{A_1}=\eta e^{-\frac{d_2-d_1}{\xi_{AF}}}$,
with $d_1=a$ and $d_2=\sqrt{5}a$ being the distance between two
holes when first and second neighbours sites are occupied (red and
blue dots in Fig. \ref{Fig_Pairon}), for each doping value, the
correction to pure $d$-wave symmetry can be found.

The doping dependence of the antiferromagnetic correlation length has been deduced from neutron experiments and is close to $1/\sqrt{p}$ \cite{PRB_Birgeneau1988,PRB_Thurston1989}. We take
\begin{equation}
\xi_{AF}(p)\propto\frac{a}{\sqrt{p-p_{AF}}},
\label{xi_AF mod}
\end{equation}
where $a$ is the lattice parameter. The parameter $p_{AF}$ has been
introduced in order to describe more precisely the data of
Ref.\,\cite{PRB_Thurston1989} than with the $1/\sqrt{p}$ dependence.
It accounts for the divergence due to the long-range order onset at
the AF transition for $p=p_{AF}\approx$ 0.02. The values of $\eta$
corresponding to each doping value is given in table
\ref{param_table}.

As indicated in the table, $\xi_{AF}(p)$ varies between
$\sim$3\,$a$, and $\sim$6\,$a$ in the entire SC doping range. Since
$a\sim$3.8\,\AA, it explains why the third and fourth nearest
neighbours do not have a significant contribution. This leaves only
the second neighbour contribution (with the coefficient $A_2$) to
fit the experimental curves of $\Delta(\theta)$. The precise fits to
the experimental ARPES data can be done with either using the
$\alpha$ parameter, and Eq.\,{\ref{formula_WS}}, or the ratio
$A_2/A_1$, using Eq.\,{\ref{2nd_neighbor}.

Both the $\alpha$ parameter and the ratio $A_2/A_1$,
representing the extension of the pairon wavefunction to the second
neighbors, vary continuously and monotonically with doping which is
a strong indication that $\Delta_N$ and $\Delta_{AN}$ have a single
origin. The evolution of the shape of $\Delta(\theta)$ as a
function of doping can thus be coherently interpreted as a direct consequence of the spatial
extension of pairons, which follows $\xi_{AF}(p)$ as a function of
hole concentration as predicted in \cite{EPL_Sacks2017}. However it should be stressed that another microscopic model, such as resonant valence bond theory \cite{Sci_Anderson1987}, which would predict a wavefunction with a similar symmetry cannot be completely ruled out.

As shown in the phase diagram of Fig. \ref{Fig_Phase_diag}, the
nodal gap has a dome shape but does not follow T$_c$.  As mentionned
before, this behavior demonstrates that the nodal gap is not the
order parameter. Examining its shape, it is tempting to extrapolate
the behavior of $\Delta_N$ for lower doping values. Its finite value
at $p=p_{min}$ would suggest that a non-zero nodal gap persists in
the small doping range $p_{AF}<p<p_{min}$ between the long range AF
order Mott transition and the superconducting state where the sample
is in an insulating state with short-range AF correlations
\cite{PRB_Takagi1989}.

This hypothesis is in agreement with the findings of Chatterjee et
al.\,\cite{NatPhys_Chatterjee2010} who measured the ARPES spectra
for a strongly underdoped non-superconducting sample of
Bi$_2$Sr$_2$CaCu$_2$O$_{8+\delta}$ ($p\approx$ 0.04), and found a
gap with nodes, as in the SC state. While the absence of deviation
from $d$-wave symmetry in this sample is not completely understood,
the presence of a gap suggests the existence of non-condensed
pairons in this system. Compatible with this discussion is the
recent discovery of a gap with a similar angular dependence as in
cuprates in electron-doped Mott insulator Sr$_2$IrO$_4$
\cite{PRL_delaTorre2015,NatPhys_Kim2016}. In both cases, it is
remarkable that the presence of a gap without superconducting
coherence may reveal an incoherent gas of pairons in the insulating
phase.

\vskip 2mm {\it Conclusion} \vskip 2mm

In this article, we have shown that the deviation of the superconducting gap from pure $d$-wave symmetry observed in cuprates by ARPES can be interpretated as the result of the existence of hole pairs, or {\it pairons}, in their local antiferromagnetic background.  In the overdoped regime, the gap has a pure $d$-wave symmetry, due to the small extension of the pairon wavefunction, restricted to first nearest neighbours. In the underdoped regime, the pairon wavefunction extends beyond first nearest neighbours, on the scale of the antiferromagnetic coherence length, which reduces the gap amplitude near the nodal direction.

The model reproduces quantitatively the measured angular dependence of the gap and its doping dependence. Furthermore it reconciles some divergences in the interpretation of ARPES data in the litterature. One concludes that the nodal and antinodal gaps originate from the same phenomenon, both intimately linked to the superconducting phase transition.

\end{document}